\documentclass{emulateapj}
\usepackage{apjfonts}

\usepackage{mathrsfs}
\usepackage{CJK}

\usepackage{amsmath}

\usepackage{graphics}

\slugcomment{Received 2016 September 17; accepted 2016 October 13}

\shorttitle{On the radio dichotomy of AGN}

\shortauthors{X. Cao}

\begin{document}

\title{On the radio dichotomy of active galactic nuclei}

\author{Xinwu Cao\altaffilmark{1,2}}
\altaffiltext{1}{SHAO-XMU Joint Center for Astrophysics, Shanghai
Astronomical Observatory, Chinese Academy of Sciences, 80 Nandan
Road, Shanghai, 200030, China; cxw@shao.ac.cn} \altaffiltext{2}{Key
Laboratory of Radio Astronomy, Chinese Academy of Sciences, 210008
Nanjing, China}

\begin{abstract}
It is still a mystery why only a small fraction of active galactic
nuclei (AGNs) contain relativistic jets. A strong magnetic field is
a necessary ingredient for jet formation, however, the advection of
the external field in a geometrically thin disk is inefficient. Gas
with a small angular velocity may fall from the Bondi radius $R_{\rm
B}$ nearly freely to the circularization radius $R_{\rm c}$, and a
thin accretion disk is formed within $R_{\rm c}$. We suggest that
the external magnetic field is substantially enhanced in this
region, and the magnetic field at $R_{\rm c}$ can be sufficiently
strong to drive outflows from the disk if the angular velocity of
the gas is low at $R_{\rm B}$. The magnetic field is efficiently
dragged in the disk, because most angular momentum of the disk is
removed by the outflows that lead to a significantly high radial
velocity. The strong magnetic field formed in this way may
accelerate jets in the region near the black hole, either by the
Blandford-Payne or/and Blandford-Znajek mechanisms. We suggest that
the radio dichotomy of AGNs predominantly originates from the
angular velocity of the circumnuclear gas. An AGN will appear as a
radio-loud (RL) one if the angular velocity of the circumnuclear gas
is lower than a critical value at the Bondi radius, otherwise, it
will appear as a radio-quiet (RQ) AGN. This is supported by the
observations that RL nuclei are invariably hosted by core galaxies.
Our model suggests that the mass growth of the black holes in RL
quasars is much faster than that in RQ quasars with the same
luminosity, which is consistent with the fact that the massive black
holes in RL quasars are systematically a few times heavier than
those in their RQ counterparts.
\end{abstract}

\keywords{accretion, accretion disks, black hole physics, magnetic
fields, galaxies: active, galaxies: jets}


\section{Introduction}\label{intro}

Active galactic nuclei (AGNs) can be divided into two categories,
i.e., radio-loud (RL) and radio-quiet (RQ) AGNs, according to their
ratios of the radio emission to optical emission. Radio emission
from RL AGNs is predominantly from the jets, and it is still a
mystery why only a small fraction (about one tenth) of AGNs exhibit
relativistic jets, while their appearance is quite similar to the RQ
counterparts in almost all wavebands except radio wavebands
\citep*[][]{1989AJ.....98.1195K,1999AJ....118.1169X,2003MNRAS.341..993C,2012ApJ...759...30B}.

The Blandford-Znajek (BZ) and Blandford-Payne (BP) mechanisms are
the most favored models of jet formation in AGNs
\citep*[][]{1977MNRAS.179..433B,1982MNRAS.199..883B}. The power of
jets is extracted from the accretion disk or the rotating black hole
by the large scale magnetic field. In the BZ mechanism, energy and
angular momentum are extracted from a rotating black hole by open
magnetic field lines, while the magnetic field threading the disk
accelerates a small fraction of the gas in the disk to form the
jets. Rapidly rotating black holes are required in RL AGNs if the BZ
mechanism is responsible for the jet formation, and therefore the
black hole spin is regarded as an intrinsic difference between RL
and RQ AGNs
\citep*[][]{1995ApJ...438...62W,2005MNRAS.357.1155Y,2005ApJ...625...72S,2007ApJ...658..815S,2008NewAR..51..891S,2010ApJ...711...50T}.
However, it is still in debate the relative importance of the BP and
BZ mechanisms
\citep*[][]{1999ApJ...512..100L,2002MNRAS.332..999C,2003ApJ...599..147C,2010MNRAS.406.1425F,2011MNRAS.418L..79T,2012MNRAS.419L..69N}.
Although we do not know which mechanism is dominant in the jet
formation of AGNs, a strong large-scale magnetic field near the
black hole is necessary for the jets in RL AGNs either with the BP
or BZ mechanisms. It is reasonable to consider the strong magnetic
field as a crucial factor causing the radio dichotomy in AGNs
\citep*[][]{2013ApJ...764L..24S}, though the black hole spin may
also play an important role in the BZ scenario. However, the
numerical simulations show that almost all massive black holes will
soon be spun up to rapidly spinning holes by accreting the gas in
the disks \citep*[][]{2007ApJ...667..704V}, though this may be
alleviated if the chaotic accretion is assumed in AGNs
\citep*[][]{2008MNRAS.385.1621K,2012ApJ...749..187L,2013ApJ...775...94V}.
This may imply that the radio dichotomy is not solely caused by the
black hole spin, instead, the radio dichotomy of AGNs may probably
originate from the strong magnetic field near the black hole, i.e.,
an AGN appears as an RL AGN only if a strong magnetic field is
present to drive relativistic jets from the region near a black
hole.

It is still unclear how the strong large scale magnetic field is
formed in the accretion disk. The external weak large-scale poloidal
field (e.g., the field threading the interstellar medium) is
suggested to be dragged inward by the accretion plasma, which is
balanced by the magnetic diffusion in the disk for a steady field
case
\citep*[][]{1974Ap&SS..28...45B,1976Ap&SS..42..401B,1989ASSL..156...99V,1994MNRAS.267..235L,2001ApJ...553..158O}.
This means that the configuration of the magnetic field dragged by
the disk is predominantly determined by the magnetic diffusivity and
the radial velocity of gas in the disk. In a conventional turbulent
accretion disk, its radial velocity is mainly regulated by the
kinematic viscosity $\nu$, and the advection of the field in the
disk is sensitive to the magnetic Prandtl number $P_{\rm
m}=\eta/\nu$, where $\eta$ is the magnetic diffusivity. It was
suggested that the magnetic Prandtl number should be around unity,
either based on the simple estimate of the order of magnitude
\citep*[][]{parker1979} or the numerical simulations
\citep*[e.g.,][]{2003A&A...411..321Y,2009A&A...504..309L,2009A&A...507...19F,2009ApJ...697.1901G}.
It was found that the field can hardly be dragged inward by a thin
disk ($H/R\ll 1$) because of its small radial velocity. The magnetic
diffusion timescale is about the same as the viscous timescale in a
steady thin disk, and the field in the inner region of the disk is
not much stronger than the external weak field
\citep*[][]{1994MNRAS.267..235L}. In order to solve this problem, a
few mechanisms were suggested to alleviate the difficulty of field
advection in the thin disks
\citep*[][]{2005ApJ...629..960S,2009ApJ...701..885L,2012MNRAS.424.2097G,2013MNRAS.430..822G,2013ApJ...765..149C}.
\citet{2013ApJ...765..149C} suggested that the most angular momentum
of the gas in the thin disk can be removed by the magnetically
driven outflows, and the radial velocity of the disk is
significantly increased. In this case, the external field can be
advected efficiently by the disk with magnetic outflows.

The jets are driven by the strong magnetic field in the region near
the black hole, either by the BP or/and BZ mechanisms. On the
assumption that an AGN will appear as an RL AGN only if a strong
magnetic field is formed near the black hole, we develop a model for
the origin of the radio dichotomy of AGNs. The model is described in
Section \ref{model}, and we put the results and discussion in
Sections \ref{results} and \ref{discussion}. The final section
contains a brief summary.

\vskip 1cm

\section{Model}\label{model}

The gas falls almost freely toward the black hole, if the angular
momentum of the gas is significantly lower than the Keplerian value
at the Bondi radius. The angular momentum of the gas is roughly
conserved until it approaches the circularization radius
\citep*[e.g.,][but also see Bu \& Yuan 2014]{2001ApJ...553..146M}.
The external weak magnetic field is dragged in by the gas in the
region between the Bondi radius and the circularization radius, due
to the field flux freezing, and the field threading the gas is
substantially enhanced at the circularization radius. An optically
thick, geometrically thin accretion disk is formed with the
circularization radius, if the gas is supplied at an appropriate
rate. The angular momentum of the gas is removed by the turbulence
in the accretion disk.

An effective magnetic diffusivity, corresponding to a magnetic
Prandtl number of the order of unity, is caused by the turbulence in
the disk. The magnetic field advection in a thin accretion disk is
quite inefficient due to magnetic diffusion in such a turbulent
disk, because the radial velocity of a thin disk is low. It was
suggested that its radial velocity is significantly increased due to
the presence of the outflows, if the angular momentum of the disk is
removed predominantly by the magnetically driven outflows
\citep*[see][for the details]{2013ApJ...765..149C}. The field in
such a disk with outflows is therefore efficiently advected toward
the black hole. The strong magnetic field formed in this way may
accelerate relativistic jets in the inner region of the disk near
the black hole, either by the BP or/and BZ mechanisms. The object
containing such an accretion disk with magnetic outflows may appear
as an RL AGN, otherwise, it may be an RQ AGN.

\subsection{Accretion of the gas in the circumnuclear region of the
galaxy}\label{accretion}

The Bondi radius and the Bondi accretion rate can be calculated by
the properties of the circumnuclear gas,
\begin{equation}
R_{\rm B}={\frac {2GM_{\rm bh}}{c_{\rm s}^2}}, \label{r_b}
\end{equation}
and
\begin{equation}
\dot{M}=4\pi\lambda (GM_{\rm bh})^2c_{\rm s}^{-3}\rho, \label{mdot}
\end{equation}
where the sound speed $c_{\rm s}=(\gamma kT/\bar{\mu} m_{\rm
p})^{1/2}$, $\gamma=5/3$, and $\lambda=0.25$ are adopted
\citep*[e.g.,][]{2006MNRAS.372...21A}. For the gas rotating with a
small angular velocity $\Omega_{\rm B}$ ($\Omega_{\rm B}\ll
\Omega_{\rm K}$, and $\Omega_{\rm K}$ is the Keplerian velocity) at
the Bondi radius, the gas falls almost freely to the black hole
without loosing its angular momentum till the circularization
radius. An accretion disk is formed within the circularization
radius $R_{\rm c}$, which means that it roughly corresponds to the
outer radius of the disk, i.e., $R_{\rm out}\simeq R_{\rm c}$. At
the circularization radius $R_{\rm c}$, the gas is rotating at the
Keplerian velocity. This leads to
\begin{equation}
R_{\rm B}^2\Omega_{\rm B}=R_{\rm c}^2\Omega_{\rm K}(R_{\rm
c}),\label{r_c}
\end{equation}
i.e.,
\begin{equation}
R_{\rm c}=R_{\rm B}\tilde{\Omega}_{\rm B}^2, \label{r_c2}
\end{equation}
where $\tilde{\Omega}_{\rm B}=\Omega_{\rm B}/\Omega_{\rm K}(R_{\rm
B})$. Suppose a weak vertical magnetic field $B_{\rm ext}$ threading
the gas at the Bondi radius, we can estimate the field strength at
the circularization radius as
\begin{equation}
B(R_{\rm c})\simeq \left({\frac {R_{\rm B}}{R_{\rm
c}}}\right)^2B_{\rm ext}. \label{b_c}
\end{equation}

\subsection{Advection of the magnetic field in the thin accretion
disk}\label{adv_b}

As the advection of the field in a thin accretion disk is quite
inefficient, we consider the field advection in the disk
predominantly driven by the magnetic outflows in this section
\citep*[see][for the details]{2013ApJ...765..149C}. If the magnetic
field is sufficiently strong in the disk, a fraction of the gas at
the disk surface (or the hot corona above the disk) may be
accelerated into the outflows by the field lines threading the
rotating disk. Such outflows may carry a large amount of the angular
momentum of the gas in the disk, which may alter the disk structure
substantially \citep*[][Cao \& Lai in
preparation]{2013ApJ...765..149C,2014ApJ...788...71L,2016ApJ...817...71C}.
The radial velocity of the disk is significantly increased due to
the presence of the outflows. In this case, the radial velocity of
the disk with magnetic outflows is
\begin{equation}
V_R\simeq V_{R,\rm vis}+V_{R,\rm m},\label{v_r}
\end{equation}
where the first term is due to the conventional turbulence in the
disk, $V_R=-\alpha c_{\rm s}H/R$, and the second term is contributed
by the outflows,
\begin{equation}
V_{R,\rm m}=-{\frac {2T_{\rm m}}{R\Sigma\Omega}}.\label{v_rm}
\end{equation}
For an accretion disk with magnetically driven outflows, the
magnetic torque exerted by the outflows in unit area of the disk
surface is
\begin{equation}
T_{\rm m}={\frac {B_zB_{\phi}^{\rm s}}{2\pi}}R={\frac {\xi_\phi
 B_z^2}{2\pi}}R,\label{t_m}
\end{equation}
where $B_{\phi}^{\rm s}$ is the azimuthal component of the large
scale magnetic field at the disk surface, and
$\xi_\phi=B_{\phi}^{\rm s}/B_z$. We use a parameter $f_{\rm m}$
($f_{\rm m}=V_{R,\rm m}/V_{R,\rm vis}$) to describe the relative
importance of the outflows on the radial velocity of the disk,
\begin{displaymath}
V_R=\left(1+f_{\rm m}\right)V_{R,\rm vis}=\left(1+{\frac {1}{f_{\rm
m}}}\right)V_{R,\rm m}~~~~~~~~~~~~~~~~~~~~~~~~~~~
\end{displaymath}
\begin{equation}
=-\left(1+{\frac {1}{f_{\rm m}}}\right){\frac {2T_{\rm
m}}{R\Sigma\Omega}}=-\left(1+{\frac {1}{f_{\rm m}}}\right){\frac
{B_z B_\phi^{\rm s}}{\pi\Sigma\Omega}}.\label{v_r2}
\end{equation}
The value of $f_{\rm m}$ is predominantly determined by the
properties of the magnetically driven outflows.

The mass accretion rate of the disk at the outer radius ($R_{\rm
out}=R_{\rm c}$) is
\begin{equation}
\dot{M}=-2\pi R_{\rm out}\Sigma(R_{\rm out})V_R(R_{\rm
out})=2\left(1+{\frac {1}{f_{\rm m}}}\right)\xi_\phi R_{\rm
out}B_z^2\Omega^{-1}.\label{mdot2}
\end{equation}
Substituting Equations (\ref{mdot}), (\ref{r_c2}) and (\ref{b_c})
into Equation (\ref{mdot2}), we obtain
\begin{displaymath}
\tilde{\Omega}_{\rm B}=0.336\xi_\phi^{2/3}\left(1+{\frac {1}{f_{\rm
m}}}\right)^{1/3}~~~~~~~~~~~~~~~~~~~~~~~~~~~~~~~~~~~~~~~~~~~~~~
\end{displaymath}
\begin{equation}
\times\left({\frac {T_{\rm B}}{\rm keV}}\right)^{-5/6}\left({\frac
{B_{\rm ext}}{\rm mGauss}}\right)^{2/3} \left({\frac {M_{\rm
bh}}{10^8M_\odot}}\right)^{1/3}\dot{m}^{-1/3}, \label{omega_b}
\end{equation}
where $T_{\rm B}$ is the gas temperature at $R_{\rm B}$, and the
dimensionless mass accretion rate $\dot{m}$ is defined as
\begin{equation}
\dot{m}={\frac {\dot{M}}{\dot{M}_{\rm Edd}}}, ~~~~~\dot{M}_{\rm
Edd}=1.39\times 10^{18}\left({\frac {M_{\rm
bh}}{M_\odot}}\right)~{\rm g~s}^{-1}. \label{mdot3}
\end{equation}
As the ratio $\xi_\phi$ is in general required to be $\la 1$
\citep*[see the detailed discussion in][]{1999ApJ...512..100L}, we
derive the first condition for efficient field advection in an
accretion disk with magnetic outflows as
\begin{displaymath}
\tilde{\Omega}_{\rm B}<0.336\left(1+{\frac {1}{f_{\rm
m}}}\right)^{1/3}~~~~~~~~~~~~~~~~~~~~~~~~~~~~~~~~~~~~~~~~~~~~~~
\end{displaymath}
\begin{equation}
\times \left({\frac {T_{\rm B}}{\rm keV}}\right)^{-5/6}\left({\frac
{B_{\rm ext}}{\rm mGauss}}\right)^{2/3}\left({\frac {M_{\rm
bh}}{10^8M_\odot}}\right)^{1/3}\dot{m}^{-1/3}, \label{omega_b1}
\end{equation}

The radial advection timescale of the magnetic field in the disk is
\begin{equation}
\tau_{\rm adv}\sim {\frac {R}{|V_R|}}. \label{tau_adv}
\end{equation}
The magnetic diffusion timescale is
\begin{equation}
\tau_{\rm dif}\sim {\frac {RH\kappa_0}{\eta}}, \label{tau_dif}
\end{equation}
where $\kappa_0=B_z/B_r^{\rm s}$ at the disk surface, and $\eta$ is
the magnetic diffusivity. For a steady magnetic field, the
inclination of the field line at the disk surface can be estimated
by equating the radial advection timescale with the magnetic
diffusion timescale, i.e., $\tau_{\rm adv}=\tau_{\rm dif}$
\citep*[][]{1994MNRAS.267..235L}. This leads to
\begin{equation}
|V_R|={\frac {\alpha P_{\rm m}c_{\rm s}}{\kappa_0}},\label{v_r4}
\end{equation}
where the magnetic Prandtl number $P_{\rm m}=\eta/\nu$, and the
viscosity $\nu=\alpha c_{\rm s}H$. Comparing Equation (\ref{v_r2})
with Equation (\ref{v_r4}), we have
\begin{equation}
\kappa_0={\frac {1}{1+f_{\rm m}}}{\frac {P_{\rm
m}}{\tilde{H}}},\label{kappa_0}
\end{equation}
where the relative disk thickness $\tilde{H}=H/R$. The cold gas can
be accelerated by the magnetic field from the midplane of a
Keplerian disk if the field line is inclined at $\la 60^\circ$
($\kappa_0\la \sqrt{3}$) with respect to the disk surface
\citep*[][]{1982MNRAS.199..883B}. The inclination angle could be a
bit higher than $60^\circ$ for the hot gas accelerated by the
magnetic field with the help of the thermal pressure or/and the
radiation force of the disk
\citep*[][]{1994A&A...287...80C,2012MNRAS.426.2813C,2014ApJ...783...51C}.
For a disk without magnetic outflows, $\kappa_0=P_{\rm
m}/\tilde{H}=\tilde{H}^{-1}$ if $P_{\rm m}=1$. It means that the gas
cannot be magnetically accelerated by the magnetic field formed in a
thin, viscously driven accretion disk, because $\kappa_0\sim P_{\rm
m}/\tilde{H}\gg 1$ for a thin disk, i.e., the field lines are nearly
vertically threading the disk. For a thin accretion disk
predominantly driven by the outflows with $\tilde{H}=0.1$,
$\kappa_0\simeq 1.7$ if $f_{\rm m}=5$ for a typical value of $P_{\rm
m}=1$. This indicates that a strong magnetic field can be formed in
a thin disk if most of its angular momentum is removed by the
outflows.

The gravitational energy dissipation rate from the unit area of the
disk is
\begin{equation}
Q_+={\frac {1}{2}}\nu\Sigma\left(R{\frac {d\Omega}{dR}}\right)^2,
\label{q_plus}
\end{equation}
and the mass accretion rate is
\begin{equation}
\dot{M}=-2\pi R\Sigma V_R=-4\pi R\left(R{\frac
{d\Omega}{dR}}\right)^{-2}\nu^{-1}V_{R,\rm vis}(1+f_{\rm m})Q_+,
\label{mdot4}
\end{equation}
where Equation (\ref{v_r2}) is used. For a fixed $Q_+$ (disk
luminosity), the mass accretion rate of an accretion disk with
magnetic outflows is $f_{\rm m}$ times higher than that of a
conventional viscously driven accretion disk, because most of the
gravitational power of the gas in the disk is tapped into the
outflows by the magnetic field.

In the outer region of the disk, the local gravity of the disk may
dominate over the gravity of the black hole. In the region beyond a
critical radius $R_{\rm T}$, it may become a clumpy disk due to the
gravitational instability
\citep*[][]{1994A&A...290...19H,2001A&A...372...50C}. The critical
radius $R_{\rm T}$ can be estimated with the Toomre parameter
\citep*[][]{1964ApJ...139.1217T,1965MNRAS.130...97G},
\begin{equation}
Q_{\rm T}={\frac {\Omega_{\rm K}c_{\rm s,d}}{2\pi G\rho_{\rm d}
H}}=1.
 \label{t_parameter}
\end{equation}
The local structure of a disk accreting at a mass rate $\dot{M}$
with magnetically driven outflows, is the same as a standard thin
disk accreting at $\dot{M}/(1+f_{\rm m})$. The critical radius
$R_{\rm T}$ can be calculated with the standard accretion disk model
when the values of the parameter $f_{\rm m}$ and other disk
parameters are specified.

The accretion of the clumpy disk may be driven by the collisions of
the clumpies, which is similar to the case of dust torus in AGNs
\citep*[][]{1988ApJ...329..702K}. The large scale magnetic field
threading the clumps may not be able to survive through repeatedly
collisions. A clumpy disk region would be an obstacle for
accumulation of external magnetic field. To avoid such a clumpy
region in the disk, one requires $R_{\rm c}<R_{\rm T}$, i.e.,
\begin{equation}
\tilde{\Omega}_{\rm B}\leq \left({\frac {R_{\rm T}}{R_{\rm
B}}}\right)^{1/2}=0.0017 r_{\rm T}^{1/2}\left({\frac {kT_{\rm
B}}{1~\rm keV}}\right)^{1/2}, \label{omega_b2}
\end{equation}
where
\begin{equation}
r_{\rm T}={\frac {R_{\rm T}}{R_{\rm S}}}, ~~~~R_{\rm S}={\frac
{2GM_{\rm bh}}{c^2}}, \label{r_t}
\end{equation}
and Equation (\ref{r_b}) is used. This is the second condition for
efficient field advection in an accretion disk with magnetic
outflows.



 \figurenum{1}
\centerline{\includegraphics[angle=0,width=7.5cm]{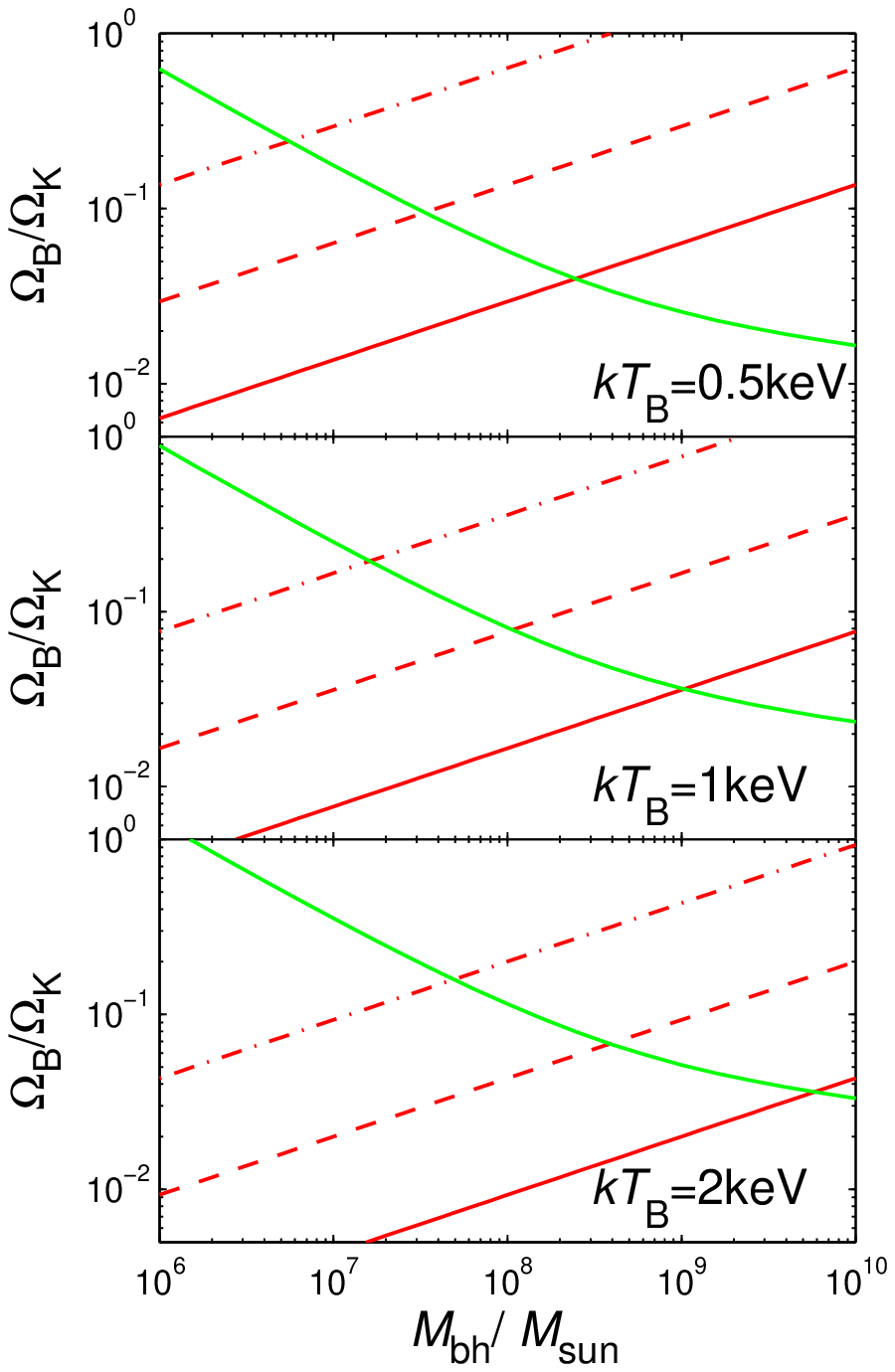}}
\figcaption{The critical angular velocities of the gas at the Bondi
radius as functions of the black hole mass. The external field can
be efficiently dragged in by an accretion disk with magnetic
outflows only if the angular velocity of the gas is lower than both
of the two critical values. All the calculations are carried out
with $\dot{m}=1$. The red lines correspond to the first condition
(see Equation \ref{omega_b1}), and the green lines to the second
condition for efficient field advection in the accretion disk with
magnetic outflows (see Equation \ref{omega_b2}). The viscosity
parameter $\alpha=1$ is adopted in the calculations. The different
types of the red lines indicate the results with different values of
the external field strength at the Bondi radius (solid: 0.01~mGauss,
dashed: 0.1~mGauss, and dash-dotted: 1~mGauss).
\label{omega_mbh}}\centerline{}

\vskip 1cm

\section{Results}\label{results}

In Section \ref{adv_b}, two constraints on the angular velocity of
the gas at the Bondi radius are derived. The external field can be
efficiently advected in an accretion disk with magnetic outflows,
only if the gas with angular velocity lower than both of these two
critical values at the Bondi radius. In Figure \ref{omega_mbh}, we
plot the results varying with the black hole mass for different gas
temperatures and external field strengths. It is found that the
angular velocity $\Omega_{\rm B}$ is always required to be much
lower than the Keplerian velocity, in order to form a strong field
in the inner region of the disk, though the detailed results also
vary with the other parameters. The results varying with the mass
accretion rate are plotted in Figure \ref{omega_mdottb1kev}. As
discussed in Section \ref{adv_b}, the model parameter $f_{\rm m}=5$
is adopted in all the calculations.


\figurenum{2}
\centerline{\includegraphics[angle=0,width=7.5cm]{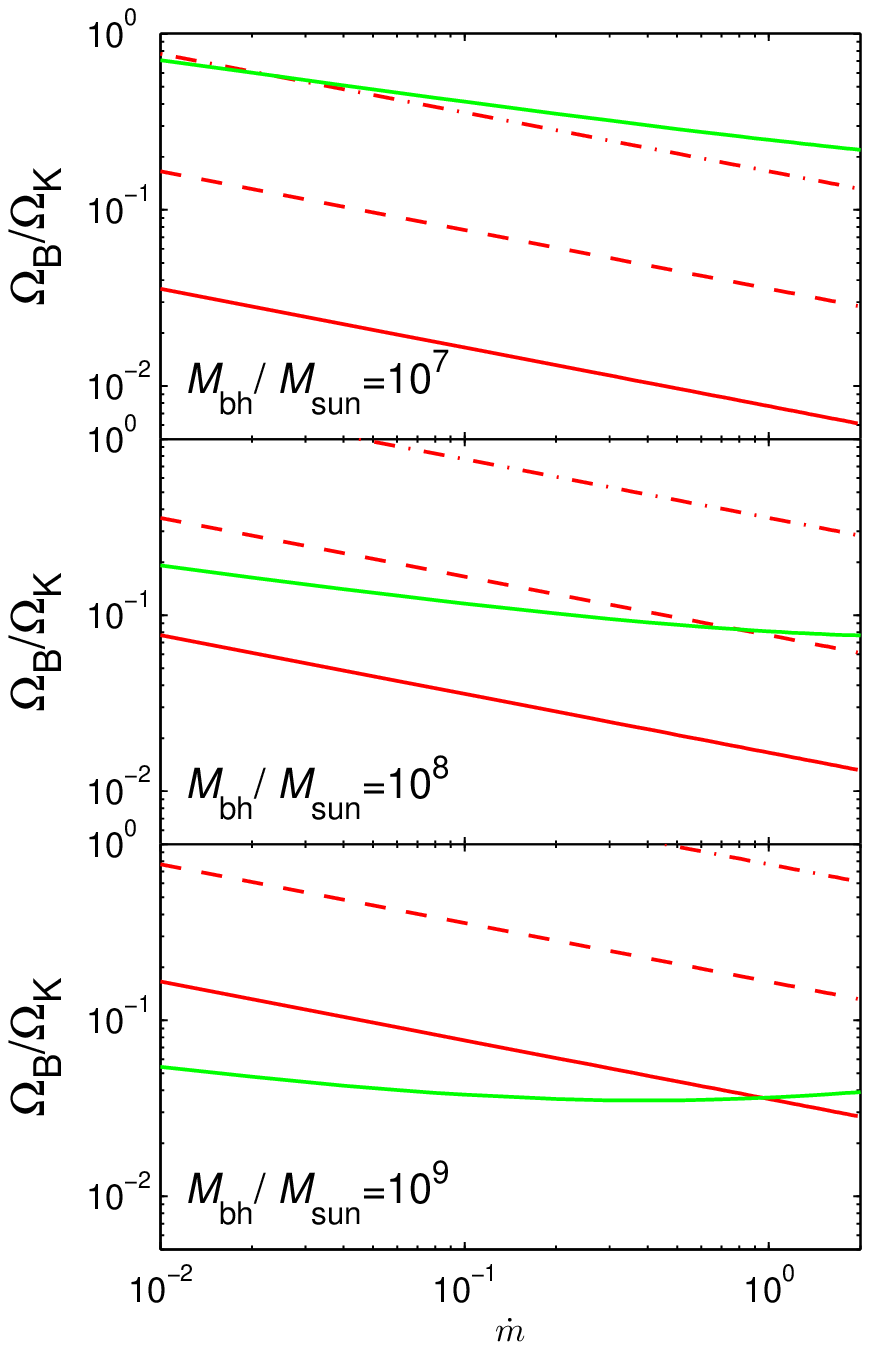}}
\figcaption{Same as Figure \ref{omega_mbh},
 but for the results as functions of the mass accretion rate at the Bondi
 radius. The circumnuclear gas temperature $kT_{\rm B}=1~{\rm keV}$
 is adopted in the calculations.
   \label{omega_mdottb1kev}}\centerline{}

\vskip 1cm

\section{Discussion}\label{discussion}

There are two conditions that should be satisfied for efficient
advection of the field in the thin disk with magnetic outflows. The
first condition is a sufficiently strong magnetic field at the outer
radius (circularization radius) of the disk (Equation
\ref{omega_b1}), and the second one is the disk not being suffered
from the gravitational instability, i.e., the local gravity of the
disk should not dominate over the gravity of the black hole
(Equation \ref{omega_b2}).

We find that RL AGNs are mainly constrained by the second condition
for the AGNs with most massive black holes ($\ga 10^9 {M}_\odot$).
For the moderate or small black holes, RL AGNs are predominantly
controlled by the first condition, unless for a strong external
magnetic field with $B_{\rm ext}\ga 0.1~{\rm mGauss}$ is present
(see Figure \ref{omega_mdottb1kev}). In this case, the critical
angular velocity of the gas at the Bondi radius, below which an AGN
may appears as a RL AGN, increases with increasing black hole mass
if all the other parameters are fixed.

Typical magnetic field strengths of galaxy cluster atmospheres are
at the order of $\sim\mu{\rm Gauss}$ \citep*[see][and the references
therein]{2002ARA&A..40..319C}, and the field strengths could be
stronger for the ISM in galaxies
\citep*[][]{2006ApJ...645..186T,2012A&A...547A..56D}. In the central
region of our galaxy, the field strength of the gas can be as high
as $\sim{\rm mGauss}$ \citep*[][]{2007A&A...464..609H}. A weaker
magnetic field requires the gas with a lower angular momentum at the
Bondi radius for RL AGNs, because the circularization radius
decreases with decreasing angular momentum of the gas, which
increases the amplification of the external magnetic field in the
region between the Bondi radius and the circularization radius.
Using \textit{Chandra} X-ray observations of nine nearby radio
galaxies, \citet{2006MNRAS.372...21A} measured the temperatures of
the gas at the Bondi radius of these galaxies, which are in the
range of $\sim 0.5-1.3$~keV. We find that our main results are
insensitive to the gas temperature (see Figure \ref{omega_mbh}).

The radio properties of AGNs are found to be linked to their host
galaxies
\citep*[e.g.,][]{2005A&A...439..487D,2006A&A...447...97B,2006A&A...453...27C}.
\citet{2005A&A...440...73C} compiled a sample of AGNs in nearby
early-type galaxies with available archival HST images. The sources
are classified with HST images into ``core" and ``power-law"
galaxies, and they found that the core galaxies invariably host a
radio-loud nucleus
\citep*[][]{2006A&A...447...97B,2006A&A...453...27C}. The core
galaxies are slowly rotating and have boxy isophotes, while the
power-law galaxies rotate rapidly and are disky
\citep*[][]{2005A&A...440...73C}. This supports the model suggested
in this paper that the source accreting the gas with a low angular
velocity may preferentially appear as an RL AGN.

The mass accretion rate is about $f_{\rm m}$ times higher than a
conventional accretion disk with the same luminosity (see Equation
\ref{mdot4}), which implies that the mass growth of the black holes
in RL AGNs is much faster than that in RQ AGNs with the same
luminosity. This is consistent with the fact that the massive black
holes in RL quasars are systematically a few times heavier than
those in their RQ counterparts
\citep*[][]{2000ApJ...543L.111L,2004MNRAS.353L..45M}. As the black
hole acquires the angular momentum of the gas in the disk, the hole
is spun up through accretion simultaneously. It may imply that most
massive black holes in RL AGNs are rotating rapidly, and the BZ
mechanism may also play an important role in RL AGNs.

Due to the mass loss in the outflows, the mass accretion rate
decreases with decreasing radius in the disk. The mass loss rate in
the outflows is governed by the magnetic field
configuration/strength and the disk properties (e.g., disk
temperature and density). The present analysis focuses on the
necessary conditions for an accretion disk-outflow system, and the
minimal field strength at the outer radius of the disk is derived
for launching strong outflows. The properties of the outflow can be
derived with the magnetic outflow solution, if suitable boundary
conditions are provided
\citep*[e.g.,][]{1994A&A...287...80C,2014ApJ...783...51C}, which is
beyond the scope of this work.

Our model implies that RL AGNs are closely associated with the
magnetic outflows. The blueshifted Fe K absorption lines in the
X-ray spectra of broad-line radio galaxies have been observed with
\textit{Suzaku}, which indicates the ultra-fast outflows co-exist
with the relativistic jets in these sources
\citep*[][]{2010ApJ...719..700T}. This seems to be consistent with
our model.

Advection dominated accretion flows (ADAFs) may probably surround
the black holes in low-luminosity AGNs \citep*[][see Yuan \& Narayan
2014 for a recent review]{1994ApJ...428L..13N,1995ApJ...452..710N}.
Although it has already been shown that the external magnetic field
can be efficiently advected in an ADAF
\citep*[][]{2011ApJ...737...94C}, the field enhancement in the
region from the Bondi radius to the circularization radius provides
a stronger field to be dragged in by the ADAF. Thus, the
circumnuclear gas with a smaller angular velocity may help to form a
stronger field in the inner region of the ADAF in low-luminosity
AGNs.

{The putative dust torus is an important ingredient of the
unification model for AGN \citep*[][]{1993ARA&A..31..473A}. It is
known that the mid-IR spectra and IR-to-optical flux ratios are very
similar in RL and RQ quasars
\citep*[e.g.,][]{2011ApJS..196....2S,2016MNRAS.461.2346G}, and dust
torus may also be present in most RL quasars if not all. We
conjecture that, the hot gas feeding the black hole co-exists with
the dust clumps in a region beyond the circularization radius, and
the dynamics of the hot gas is not altered significantly by the
motion of the dust clumps in the torus. If this is the case, the
accretion of the hot gas from the Bondi radius to the black hole can
still be well described by the calculations in this paper even if a
putative dust torus is present. }

\vskip 1cm

\section{Summary}\label{summary}

The circumnuclear gas with a low angular velocity falls nearly
freely from the Bondi radius to form an accretion disk within the
circularization radius. The external magnetic field threading the
gas is strongly amplified in this region due to the field flux
freezing. For the gas with an angular velocity lower than a critical
value, the field in the disk is strong enough to drive magnetic
outflows, which carries away most of the angular momentum of the
disk. This strongly increases the radial velocity of the disk, and
therefore the field can be efficiently dragged inwards by the disk.
Relativistic jets may be driven by such a large scale magnetic field
through either the BP or BZ mechanism. In this case, the object may
appear as an RL AGN.

If the angular velocity of the circumnuclei gas is larger than a
critical value, the circularization radius (i.e., the outer radius
of the disk) becomes larger, and the field cannot be amplified to a
strong field in the disk to accelerate outflows. Thus, a
conventional turbulence driven thin disk is formed within the
circularization radius, and the advection of the field in the disk
is rather inefficient. In this case, no relativistic jet is formed
in the inner region of the disk, which corresponds to an RL AGN.

\acknowledgments I thank the referee for the helpful
comments/sugestions. This work is supported by the NSFC (grant
11233006), the Strategic Priority Research Program ¡°the Emergence
of Cosmological Structures¡± of the CAS (grant No. XDB09000000), the
CAS/SAFEA International Partnership Program for Creative Research
Teams, and Shanghai Municipality.

{}

\end{document}